\documentclass[prb,twocolumn,showpacs,preprintnumbers,amsmath,amssymb]{revtex4}

\usepackage{graphicx}
\usepackage{amssymb}
\usepackage{amsmath}
\usepackage{bm}
\usepackage{graphicx}
\usepackage{color}

\begin{document}

\title{Spin-wave velocities, density of magnetic excitations,
and NMR relaxation in ferro-pnictides}

\author{Andrew Ong}
\affiliation{School of Physics, University of New South Wales, 
Sydney 2052, Australia}
\author{G\"otz S. Uhrig \footnote{On leave from
Lehrstuhl f\"{u}r Theoretische Physik I,
Technische  Universit\"{a}t Dortmund,
 Otto-Hahn Stra\ss{}e 4, 44221 Dortmund, Germany}}
\affiliation{School of Physics, University of New South Wales, 
Sydney 2052, Australia}
\author{Oleg P. Sushkov}
\affiliation{School of Physics, University of New South Wales, 
Sydney 2052, Australia}

\begin{abstract}
We perform an analysis of the experimentally known temperature dependence
of the staggered magnetization in the antiferromagnetic phase.
This analysis allows us to put an upper limit on the unknown value
of the spin wave velocity along the stripes of equal spin direction
(spin stripes). The velocity is about ten times smaller than the velocity 
perpendicular to the spin stripes. 
The strongly anisotropic spin-wave dispersion 
implies a high density of low energy magnetic excitations.
We demonstrate that this high density strongly enhances the
$^{75}$As NMR spin-lattice  relaxation via the Raman scattering of magnons. 
We derive the polarization dependence of this relaxation channel
and find very good agreement with experimental data.
The high density of low energy magnetic excitations deduced from 
our phenomenological analysis supports the scenario that ferro-pnictides
are close to a quantum phase transition.
\end{abstract}


%74.70.-b Superconducting materials (for cuprates, see 74.72.¡Ýh) 
%75.25.+z Spin arrangements in magnetically ordered materials (including neutron and spin-polarized electron studies, synchrotron-source X-ray scattering, etc.) (for devices exploiting spin polarized transport, see 85.75.¡Ýd) 
%75.30.Ds Spin waves (for spin-wave resonance, see 76.50.+g) 
%75.50.Ee Antiferromagnetics  
%75.40.Gb Dynamic properties (dynamic susceptibility, spin waves, spin diffusion, dynamic scaling, etc.)  
%75.10.Jm Quantized spin models  
%76.60.-k Nuclear magnetic resonance and relaxation (see also 33.25.+k Nuclear resonance and relaxation in atomic and molecular physics and 82.56.-b Nuclear magnetic resonance in physical chemistry and chemical physics; for structure determination using magnetic resonance techniques, see 61.05.Qr; for biophysical applications, see 87.80.Lg)
%76.60.Es Relaxation effects 

\date{\today}

\maketitle

\section{Introduction}
One of the most important and widely discussed issues in the physics
of iron pnictide superconductors is whether these materials are strongly or 
weakly correlated. A closely related issue is the origin of magnetism in their
parent compounds. In a slightly simplistic way one can formulate the problem 
in the following way. Does the magnetism arise from itinerant electrons or is 
it due to localized electrons?
For a recent review,  see for instance Ref.\ \onlinecite{mazin09}.
The situation is different from cuprates where parent compounds are
clearly Mott insulators and hence there is no ambiguity about the origin of 
magnetism.

\begin{figure}[ht]
\includegraphics[width=0.3\textwidth,clip]{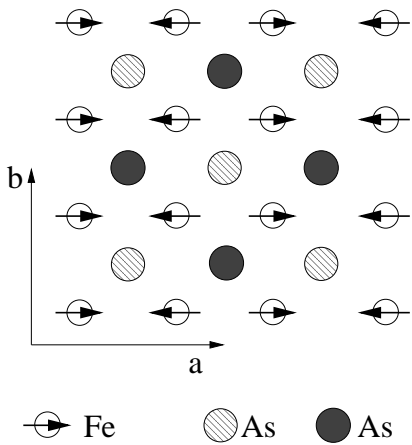}
\caption{The Fe-As plane. Fe ions are shown by open circles and As ions
are shown by filled circles. The Fe spins are shown by arrows.
The Fe ions lie exactly in the plane while As ions are out of plane 
by $\delta_c \approx \pm 1.35$\AA{} in a checkerboard pattern.
The pattern is shown by different fillings of the symbols for the As ions.}
\label{cell}
\end{figure}

In the present work, we do not address the issue of strong or weak correlations
directly. In a phenomenological way we analyse available experimental data 
on low temperature magnetic properties and determine the previously unknown 
spin-wave velocity along the spin stripes,
 by which we refer to chains of spins 
running in $b$ direction in which the spins point all in the same direction
(cf.\ Fig.\ \ref{cell}). 
This spin wave velocity turns out to be very small. 
It is by an order of magnitude smaller than the velocity
perpendicular to the spin stripes. The knowledge of the velocity is very 
important itself  because it predicts the outcome of 
future inelastic neutron scattering measurements. 
In addition, the knowledge sheds light on the issue of strong
or weak correlations.
The low velocity implies a high density of magnetic excitations and
the high density strongly supports the  strong correlation scenario based
on the vicinity to a quantum critical point~\cite{fang08,xu08,yao08,uhrig09a}.
Due to the high spectral density magnons must contribute significantly to
the NMR relaxation rate at a temperature above the spin-wave gap.
We consider this mechanism for $^{75}$As NMR spin-lattice relaxation
and find a very good agreement between theoretical results 
 and experimental data.

The magnetic long range order is firmly established in the parent compounds
LaFeAsO and Sr(Ba,Ca)Fe$_2$As$_2$
by neutron scattering  \cite{cruz08,zhao08,ewing08,mcque08},
muon spin resonance, and Moessbauer spectroscopy \cite{klaus08,luetk09}.
The neutron scattering  reveals a columnar antiferromagnetic ordering with a 
staggered magnetic moment of $(0.3-0.4)\mu_\text{B}$ in LaFeAsO and 
$(0.8-0.9)\mu_\text{B}$ in Sr(Ba,Ca)Fe$_2$As$_2$.
All the compounds are layered systems consisting of Fe-As planes.
For simplicity, we consider only the tetragonal  lattice which is formed 
by the Fe ions ignoring a small orthorhombic and even monoclinic structural 
distortion. In Fig.\ \ref{cell} we show schematically the Fe-As plane and the 
spin ordering at the Fe sites.
Along the $a$ axis the spin directions alternate whereas they are the same
along the $b$ axis. Spins also alternate along the $c$ axis which 
is orthogonal to the plane.
%%%%%%%%%%%%%%%%%
%%%%%%%%%%%%%%%%%
In our study, the lattice spacings, i.e., the distances between Fe ions, are 
$g_a\approx g_b\approx 2.79$\AA{}
and $g_c \approx 6.15$\AA.
We choose units such that all lattice spacings equal unity, $g_a\to 1$,
$g_b\to 1$, $g_c\to 1$. Note that the arsenic ions are shifted out of plane by 
$\delta c \approx \pm 1.35$\AA{}
in a checkerboard pattern shown in Fig.\ \ref{cell}.

The spin wave velocities along the $a$ and the $c$ axis as well as the 
spin wave gap at zero temperature have been measured by neutron 
scattering for SrFe$_2$As$_2$~\cite{zhao08} and for 
BaFe$_2$As$_2$~\cite{ewing08},
\begin{eqnarray}
\label{vac}
v_a&\approx& 205\text{meV} \ , \nonumber\\
v_c&\approx&  45 \text{meV} \ , \nonumber\\
\Delta(T=0)&\approx& 6.5\text{meV} \ .
\end{eqnarray}
But the spin-wave velocity along the $b$ axis, i.e., along the spin stripes, 
has  not yet been measured to our knowledge.
Note that the unit cell lattice spacings for the
122 compounds SrFe$_2$As$_2$ and 
BaFe$_2$As$_2$ are twice larger than the corresponding values of $g$, 
$a =2g_a$, $b =2g_b$, $c =2g_c$.
The standard crystallographic convention  is to set
$a \to 1$, $b\to 1$, $c\to 1$. Thus the values of the spin wave velocities 
in these standard units are twice larger than the values in our units.

\begin{figure}[ht]
\includegraphics[width=0.98\columnwidth,clip]{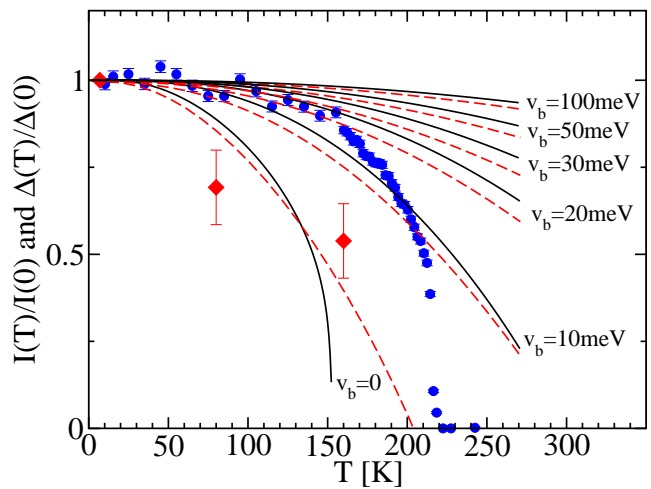}
\caption{(color online) Temperature dependence of the intensity of 
elastic neutron scattering at the antiferromagnetic superlattice 
reflection and of the spin-wave gap in SrFe$_2$As$_2$.
Points with error bars show experimental data from Ref.~\onlinecite{zhao08}.
Blue circles show the normalized neutron scattering intensity,  $I(T)/I(0)$,
and red diamonds show the normalized spin-wave gap, $\Delta(T)/\Delta(0)$.
The curves show theoretical results for the 
normalized neutron scattering intensity
for various values of the spin-wave velocity $v_b$ along the spin 
stripes. Solid black curves correspond to the first scenario for the spin-wave 
gap, Eq.\ (\ref{c1}), and dashed red curves correspond to the second 
scenario, Eq.\ (\ref{c2}).  The theory is justified
only where the deviation of $I(T)/I(0)$ from unity is small.
}
\label{exp}
\end{figure}

The N\'eel temperature for these compounds is $T_N=200-220$K.
The temperature dependence of the normalized intensity of elastic neutron 
scattering, $I(T)/I(0)$, and the temperature dependence of the normalized 
spin-wave gap, 
$\Delta(T)/\Delta(0)$ have been measured in Ref.~\onlinecite{zhao08}.
These experimental results are shown in Fig.~\ref{exp}.
%%%%%%%%%%%%%%%%%
%%%%%%%%%%%%%%%%%

We are not aware of a direct measurement of the spin magnetic susceptibility 
of SrFe$_2$As$_2$. However, the data for LaFeAsO~\cite{ning08,kling09} and 
BaFe$_2$As$_2$~\cite{ning09} show that the spin susceptibility in the AF 
ordered phase averaged over directions is about 
\begin{equation}
\label{chi}
\chi_s \approx 1\times 10^{-4}\text{emu}/(\text{mol Fe}) \ .
\end{equation}
We will use this value for SrFe$_2$As$_2$ having in mind that it might be 
by a factor $\approx 1.5-2$ off.

\section{Effective action for magnetic excitations}
To describe spin waves we use an effective model, 
the nonlinear $\sigma$-model with the  following Lagrangian
\begin{eqnarray}
\label{L}
&&L = \nonumber \\
&&\frac{\chi_\perp}{2}
\left\{\dot{\vec{n}}^2-v_a^2(\partial_a{\vec n})^2
-v_b^2(\partial_b{\vec n})^2
-v_c^2(\partial_c{\vec n})^2
+\Delta_b^2n_a^2\right\}\nonumber\\
&& = \frac{\chi_\perp}{2}
\left\{(\partial_{\mu}{\vec n})^2 +\Delta_b^2n_a^2\right\} \ ,
\end{eqnarray}
where
\begin{equation}
\partial_{\mu}=(\partial_t,\ i v_a\partial_a,\ i v_b\partial_b, \ 
i v_c\partial_c) \ ,
\end{equation}
and $\Delta_b$ is the bare spin-wave gap.
The standard constraint ${\vec n}^2=1$ is imposed.
Hereafter we set  $k_B=\hbar=1$ for simplicity.
Note that generally the bare spin-wave gap $\Delta_b$ can depend on temperature
if the temperature dependence arises from physics different from spin waves, 
say from phonons.
We stress that this phenomenological description of low energy magnetic 
excitations is valid independently of the specific mechanism for magnetism.
The description is equally valid for  magnetism caused by  itinerant electrons 
and for magnetism caused by localized electrons.
We will use (\ref{L}) below the N\'eel temperature $T_N$.
The field theory (\ref{L}) is the only possible effective  theory
that describes spin waves with dispersion
$\omega_q=\sqrt{v_a^2q_a^2+v_b^2q_b^2+v_c^2q_c^2+\Delta^2}$.
Therefore, the only important issue for the justification of (\ref{L}) is 
that there are  well-defined low-energy spin waves.
This is directly supported by experiment~\cite{zhao08,ewing08}. 
According to Ref.~\onlinecite{ewing08} the spin waves are well-defined 
up to $\omega \approx 150$meV.
This is an important piece of information, but pragmatically, for purposes of 
the present work, we only need that spin waves exists with energies 
$\omega \lessapprox T_N\approx 20$meV.

The spin-wave velocities $v_a$ and $v_c$ are known from experiment, 
see Eq.\ (\ref{vac}). The susceptibility $\chi_{\perp}$ is related to the spin 
magnetic susceptibility
(\ref{chi})
\begin{equation}
\label{chi1}
\chi_s=\chi_{\perp} \frac{2}{3}(g\mu_B)^2N_A \ ,
\end{equation}
where $g$ is the gyromagnetic ratio, $\mu_B$ is the Bohr magneton,  $N_A$ is 
the Avogadro constant, and the factor 2/3 comes from  averaging over 
orientations. Lacking any other information,
we take the standard value of the gyromagnetic ratio, $g=2$. 
Eqs.\ (\ref{chi1}) and (\ref{chi}) yield
\begin{equation}
\label{chi2}
\chi_{\perp} =1.2 \cdot 10^{-3} 1/\text{meV} .
\end{equation}
Due to the uncertainty in the values for $\chi_s$ and $g$ one 
has to face an uncertainty in the value of $\chi_{\perp}$. At worst,
we estimate the uncertainty in the value of $\chi_{\perp}$ to be
a factor 1.5-2 relative to the value given in Eq.\ (\ref{chi2}).

Assuming that the system is below the N\'eel temperature, 
${\vec n} \approx (n_a,0,0)$,  we represent the staggered magnetizaton by
\begin{equation}
{\vec n} = (n_a,n_b,n_c)=(\sqrt{1-\pi^2},  \vec{\pi}) \ ,
\end{equation}
where the field ${\vec \pi}$ is two-dimensional having 
only $b$- and $c$-components.
The static component of the staggered magnetization reads
in the first two leading orders
\begin{equation}
\label{nas}
\langle n_a\rangle\approx 1-\frac{1}{2}\langle {\vec \pi}^2\rangle \ ,
\end{equation}
where $\langle \ldots \rangle$ denotes the quantum  Gibbs expectation value.

Expanding Eq.\ (\ref{L}) in powers of ${\vec \pi}$ up to quartic terms 
(single loop corrections) we obtain the following Lagrangian
for the $\pi$-field
\begin{equation}
\label{Lpi}
L = \frac{\chi_\perp}{2}\left\{(\partial_{\mu}{\vec \pi})^2
+{\vec \pi}^2(\partial_{\mu}{\vec \pi})^2-\Delta_b^2{\vec \pi}^2\right\} \ .
\end{equation}
In order to derive the temperature dependent quadratic effective Lagrangian 
$L_T$ one has to perform a decoupling in the quartic term in (\ref{Lpi})
\begin{eqnarray}
\label{dec}
{\vec \pi}^2(\partial_{\mu}{\vec \pi})^2 &\to&
\langle{\vec \pi}^2\rangle(\partial_{\mu}{\vec \pi})^2+
{\vec \pi}^2\langle(\partial_{\mu}{\vec \pi})^2\rangle\nonumber\\
&\to&
\langle{\vec \pi}^2\rangle(\partial_{\mu}{\vec \pi})^2+
\Delta_b^2{\vec \pi}^2\langle{\vec \pi}^2\rangle \ .
\end{eqnarray}
Here we have used integration by parts and the
equation of motion in leading order
\begin{equation}
\label{eqm}
\partial_{\mu}^2{\vec \pi}+\Delta_b^2{\vec \pi}=0\ .
\end{equation}
Hence the decoupling yields the following effective Lagrangian
\begin{equation}
\label{Lt}
L_T = \frac{\chi_\perp}{2} \left\{[1+\langle {\vec \pi}^2\rangle]
(\partial_{\mu}{\vec \pi})^2
-\Delta_b^2[1-\langle {\vec \pi}^2\rangle]{\vec \pi}^2\right\} \ .
\end{equation}
After rescaling the field, 
${\vec \pi}_R=\sqrt{1+\langle {\vec \pi}^2\rangle}{\vec \pi}$, we find that
the spin-wave gap is renormalized as
\begin{equation}
\label{gg}
\Delta(T) = (1-\langle {\vec \pi}^2\rangle )\Delta_b \ .
\end{equation}

Following Fermi's Golden Rule, the
scattering intensity $I(T)$ is proportional to the square of 
the staggered magnetization $\langle n_a\rangle^2$ 
which is given by Eq.\ \eqref{nas}.
Combining this fact with  Eq.\ \eqref{gg} yields
\begin{equation}
\label{tIG}
\frac{I(T)}{I_b}=\frac{\Delta(T)}{\Delta_b}=1-\langle {\vec \pi}^2\rangle 
+{\cal O}(\langle {\vec \pi}^2\rangle^2)\ ,
\end{equation}
where $I_b$ is a bare scattering intensity without quantum fluctuations,
for further discussion see below. 
In deriving this equation we assume formally
that $\langle {\vec \pi}^2\rangle \ll 1$. We will discuss this point
in more detail below.

A standard calculation of the expectation value $\langle {\vec \pi}^2\rangle$
leads to the following  result
\begin{equation}
\label{pi2}
\langle {\vec \pi}^2\rangle = \frac{1}{\chi_{\perp}}
\int\frac{d^3q}{(2\pi)^3}\frac{1}{\omega_q}(2n(\omega_q)+1)
 \ 
\end{equation}
where 
\begin{equation}
\label{disp}
\omega_q=\sqrt{v_a^2q_a^2+v_b^2q_b^2+v_c^2q_c^2+\Delta^2(T)}
\end{equation} 
is the spin-wave dispersion and
\begin{equation}
\label{BE}
n(\omega_q)=\frac{1}{e^{\omega_q/T}-1}
\end{equation} 
is the Bose-Einstein distribution function.
The unity in the factor $(2n_q+1)$ in (\ref{pi2}) is due to 
quantum fluctuations which lead to the quantum renormalization of
the bare spin wave gap, $\Delta_b(T) \to \Delta_{qr}(T)$,
and to the concomitant renormalization of the bare scattering intensity,
$I_b \to I_{qr}(T)$. The quantitative outcome of this renormalization depends
on the high-energy cutoff
\begin{subequations}
\begin{eqnarray}
\frac{I_{qr}}{I_b} = \frac{\Delta_{qr}}{\Delta_b} = 1- \frac{1}{\chi_{\perp}}
\int\frac{d^3q}{(2\pi)^3}\frac{1}{\omega_q} .
\end{eqnarray}
\end{subequations}
In our study, we include the case that the quantum renormalized quantities
retain a temperature dependence  from a temperature dependent $\Delta_b(T)$, 
whose dependence is induced from physical effects outside
of the non-linear $\sigma$-model, for instance from structural changes
or phonons. The quantum renormalized quantities
equal the physical ones at zero temperature $\Delta(0)=\Delta_{qr}(0)$
and $I(0)=I_{qr}(0)$. Note that  $I_b$ is temperature
independent by definition.

On the present single loop level, the temperature effects
can be accounted for by
\begin{subequations}
\begin{eqnarray}
\label{tIG-renorm}
\frac{I(T)}{I_{qr}(T)} &=&\frac{\Delta(T)}{\Delta_{qr}(T)}=1-\langle {\vec \pi}^2
\rangle_\text{therm}\\
\label{pi2-renorm}
\langle {\vec \pi}^2\rangle_\text{therm}& =& \frac{1}{\chi_{\perp}}
\int_0^\infty \rho(\omega,\Delta(T)) n(\omega)d\omega
\end{eqnarray}
\end{subequations}
instead of Eqs.\ \eqref{tIG} and \eqref{pi2}.
Here we use the density of magnetic excitations $\rho(\omega,\Delta)$
which reads in three dimensions
\begin{equation}
\label{md3}
\rho(\omega,\Delta) = \frac{1}{\pi^2 v_a v_b v_c}\omega
\sqrt{\omega^2-\Delta^2}\ \Theta(\omega-\Delta) \ .
\end{equation}
The above continuum expressions in three dimensions are only valid
if $v_b$ is not very small compared to temperature, $v_b\pi \gg T$, because for 
dominating temperature the boundaries of the Brillouin
zone are felt which are not captured by the non-linear $\sigma$ model.
In the opposite limit, $v_b\pi \ll T$  one should use the
the two-dimensional density of 
\begin{equation}
\label{md2}
\rho(\omega,\Delta) = \frac{1}{\pi v_a v_c}\omega\ \Theta(\omega-\Delta) \ .
\end{equation}

We point out that for temperature independent bare gap 
$\Delta_b$ we have $I_{qr}(T)=I(0)$ and $\Delta_{qr}(T)=\Delta(0)$.
Then Eq.\ \eqref{tIG-renorm} already provides the result to be compared
with experiment. Both normalized quantities, gap and intensity,
should display the same
temperature dependence in their deviation from unity.
For this reason they are depicted in the same plot in Fig.\ \ref{exp}.
We will discuss the very different behaviour of both experimental
quantities below.

Eq.\ \eqref{tIG-renorm} does not yet provide the 
ratio $I(T)/I(0)$ given by experiment if $\Delta_b$ is temperature dependent.
 To obtain full knowledge about $I(T)/I(0)$ we have to account for the 
influence of the infrared cutoff, i.e., the gap, on the quantum renormalization. We find
\begin{subequations}
\label{IR}
\begin{eqnarray}
&& R_{qr}(T) := \frac{I_{qr}(T)}{I_{qr}(0)}\\
&& = 1-\frac{1}{2\chi_{\perp}}
\int_0^\Lambda d\omega\frac{\rho(\omega,\Delta_{qr}(T))- \rho(\omega,\Delta_{qr}(0))}
{\omega},
\qquad\
\end{eqnarray}
\end{subequations}
where we introduce a high-energy (UV) cutoff $\Lambda$ to ensure 
convergence. A realistic estimate is $\Lambda=200$meV \cite{ewing08,uhrig09a}.
Note that the difference occurring in \eqref{IR} depends only weakly, i.e.,
logarithmically on the precise value of $\Lambda$. In two dimensions, there is even no
dependence on the UV cutoff at all.

The final result is obtained by combining \eqref{tIG-renorm} and \eqref{IR}
in
\begin{equation}
\label{final}
\frac{I(T)}{I(0)} = \frac{I(T)}{I_{qr}(T)} \cdot \frac{I_{qr}(T)}{I_{qr}(0)}
= \left[1-\langle {\vec \pi}^2 \rangle_\text{therm}\right] R_{qr}(T)
\end{equation}
where $I_{qr}(0)=I(0)$ entered.

A remark on the validity of the single loop approximation is in order.
Obviously, the theoretical expressions are only valid if
the thermal renormalization remains small, i.e., $\delta I/I(0) \ll 1$, $\delta I=I(0)-I(T)$.
Therefore, one cannot rely on (\ref{final}) in the vicinity of the critical 
point. But we can rely on  (\ref{final}) at $T< 200$K where, according to
experimental data, $\delta I/I(0)< 0.3$.

\section{Value of the gyromagnetic ratio and possible orbital dynamics}

In the present work we use the standard value $g=2$
which is the spin gyromagnetic ratio.
This scenario assumes that the magnetism in the system is entirely
due to spins. But the orbital physics of iron is certainly
more complex and spin-orbit coupling plays an important role \cite{wu08}.
So values of the $g$-factor or $\underline{g}$-tensor different
from a scalar value of 2 are well possible.

The phenomenological nonlinear $\sigma$-model description is valid
independent of the origin of magnetism. Therefore our conclusions  in the 
present paper do not depend on the extent that orbital and/or
charge degrees of freedom play a major role. Only the precise
numerical estimates depend on the numerical value of $g$.
So for the present paper a better knowledge
of $g$ will affect only the numerical estimates, not the
scenario. But for microscopic considerations, for instance
the issue which value of spin is most appropriate, the
local orbital physics is of fundamental importance and 
measurements of $g$ can shed light on this issue.

We are not aware of data on the value of $g$.
Hence we would like to point out that the large value
of the spin-wave gap, $\Delta(0)=6.5$meV, gives a unique opportunity
to measure $g$ by inelastic neutron scattering.
If a magnetic field ${\cal B}$ is applied that is directed along
the $a$ axis the spin wave excitations will be shifted in energy
 according to their  projection of the angular momentum
$\pm 1$ along $a$. Therefore, the magnetic field will split the spin wave gap in 
two 
\begin{subequations}
\begin{eqnarray} 
\label{tg}
\Delta_-&=&6.5\text{meV} -g\mu_B{\cal B} \ ,
\\
\Delta_+&=&6.5\text{meV} +g\mu_B{\cal B} \ .
\end{eqnarray}
\end{subequations}
For ${\cal B}=15$Tesla and for $g=2$ the splitting takes the value
$2g\mu_B{\cal B}\approx 3.5$meV so that it should be easily observable
 in neutron spectra.

\section{Analysis of Experimental Data}

For explicit calculations we need to specify the temperature dependence of 
the quantum renormalized gap $\Delta_{qr}$. 
Since we do not have experimental knowledge about
this quantity we study two scenarios which correspond
to opposite limits. It will turn out that our
conclusions depend only weakly on which scenario is realized.

The spin wave gap is caused by spin-orbit interaction
in combination with the orthorhombic lattice deformation.
The deformation is practically temperature independent below $T_N$.
Hence the  scenario (i) assumes that the quantum renormalized spin wave gap
$\Delta_{qr}$ is temperature independent. Then the observed
spin gap acquires its temperature dependence $\Delta(T)$
solely from Eq.\ \eqref{tIG-renorm}. The zero temperature
value is fixed to
\begin{equation}
\label{c1}
\Delta_{qr}=\Delta(0)=6.5\text{meV}  \ ,
\end{equation}
 and the temperature dependence of the
physical gap $\Delta(T)$ is determined
by the self-consistent solution of Eqs.\ 
\eqref{vac}, \eqref{chi2}, \eqref{tIG-renorm},
\eqref{pi2-renorm}, and \eqref{md3} or \eqref{md2}.
However, according to Eq.\ (\ref{tIG-renorm}) 
this scenario implies the identical 
temperature dependence of the normalized
neutron intensity and of the normalized spin-wave gap.
This consequence is not supported by experiment.
According to the data from Ref.\ \onlinecite{zhao08} shown in Fig.\
\ref{exp}
the dependencies are significantly different. But it cannot
be excluded that experimental difficulties, for instance
the influence of the charge degrees of freedom, prevent the reliable 
measurement of the spin gap at finite temperature. So it is instructive to 
consider  scenario (i) as one limiting case.

In  scenario (ii) we assume that the quantum renormalized gap $\Delta_{qr}$ is
temperature dependent in precisely such a way that the
experimentally observed $\Delta(T)$ shown in Fig.\ \ref{exp}
is induced. The temperature dependence $\Delta_{bR}(T)$  may result from 
the influence of low-energy phonons. 
In this case we fit the experimental data~\cite{zhao08} for $\Delta(T)$  
by the linear function
\begin{equation}
\label{c2}
\Delta(T)=6.5\text{meV}-0.020 T \text{meV}/\text{K} \ .
\end{equation}
Then we employ  Eqs.\ (\ref{vac}), (\ref{chi2}), (\ref{tIG-renorm}),
(\ref{pi2-renorm}), \eqref{md3} or \eqref{md2}  to
calculate the neutron scattering intensity resulting
from the  phenomenological spin-wave gap (\ref{c2}).
 
The theoretical results obtained in the two scenarios 
 for the values of the spin-wave velocity along the spin stripes
$v_b=100\text{meV}, 50\text{meV}, 30\text{meV}, 20\text{meV}, 
10\text{meV}$, and $0\text{meV}$ are displayed in Fig.~\ref{exp}.
The solid black curves correspond to  scenario (i), i.e., Eq.~(\ref{c1}).
The dashed red curves correspond to  scenario (ii), i.e., Eq.~(\ref{c2}).

We emphasize that both scenarios for the temperature dependence of
the spin-wave gap yield almost coinciding curves
for the neutron scattering intensity. This can be attributed 
to the fact that by construction both scenarios are equal
at $T=0$ so that the difference between them can only be discerned
at sufficiently large temperature. But if $T\gtrapprox \Delta(T)$
the precise value of the spin wave gap does not matter anymore.

We already pointed out that the comparison between theory and 
experimental data makes sense only below 180-200K where $\delta I/I(0)$ 
is sufficiently small and the theory is quantitatively reliable.
The curves with $v_b=100$meV and $v_b=50$meV
 clearly disagree with experiment,
while the curves with $v_b=30$meV and $v_b=10$meV 
constitute upper and lower bounds to the experimental data.
The curve for $v_b=20$meV is in good agreement  with the data in the
range of its validity ($\delta I/I \ll 1$). The 3D formula (\ref{md3}) 
is at the verge of its validity for $v_b=10$meV since  $v_b\pi \approx T$.
Therefore, in Fig.~\ref{exp} we also include curves for $v_b=0$ which
are obtained using the 2D formula (\ref{md2}).
We also studied the 3D $\to$ 2D crossover empirically replacing 
$v_b^2q_b^2$ by $4v_b^2\sin^2(q_b/2)$ in Eq.\ (\ref{disp}).
We do not show the corresponding curves because they completely 
confirm the results shown in Fig.~\ref{exp}.

Thus, our conclusion is that the value of the spin-wave velocity along spin 
stripes, $v_b$ is in the range 
\begin{equation}
\label{vb}
v_b \approx 10-30\text{meV},
\end{equation}
 that means it is at least ten times smaller than  $v_a\approx 205$meV.
We stress that in essence this conclusion is based on the density of magnetic
excitations \eqref{md3}.
The number of excitations at low energies govern the thermally
induced reduction of the staggered magnetic moment.
If $v_b$ is large, the density is low implying a weak
temperature dependence of the elastic neutron scattering intensity.
If $v_b$ is small, the density is large implying a strong
temperature dependence of the elastic neutron scattering intensity.
The value of $v_b$ must be sufficiently low to produce the experimentally 
found temperature dependence of the scattering intensity.

The very low value of $v_b$ that follows from the experimental
result for $I(T)$ implies that magnetic fluctuations play
a prominent role in the ferro-pnictides. This supports
the previously proposed scenario that the ferro-pnictides
are systems close to a quantum phase transition \cite{uhrig09a}.

\section{NMR spin-lattice relaxation for $^{75}$As.}

\begin{figure}[ht]
\includegraphics[width=0.5\columnwidth,clip]{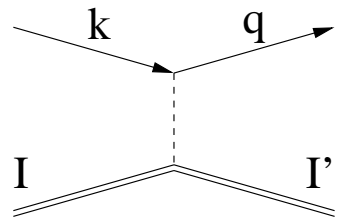}
\caption{Magnon Raman scattering on nuclear spin due to
hyperfine interaction. Solid lines denote magnons, the double lines
denote a nuclear spin. The dashed line denotes the hyperfine interaction.}
\label{hfsfig}
\end{figure}

As an additional testbed for our scenario of strong magnetic
fluctuations at low energies we study the NMR relaxation rate $T_1^{-1}$.
This relaxation is due to inelastic Raman type scattering of thermally 
excited magnons from nuclear spin, see Fig.~\ref{hfsfig}.
%%%%%%%%%%%%%%%%%
%%%%%%%%%%%%%%%%%
Clearly, this mechanism is most important at temperatures above the 
spin-wave gap. For concreteness, we consider the relaxation of 
the nuclear spin of $^{75}$As. The As ion is positioned at the top of a
 pyramid with four Fe ions at its base, see Fig.~\ref{cell}. 
The hyperfine interaction of the $^{75}$As nuclear spin $I$ with
the electronic spin $S$ on the  adjacent Fe ion is of the following 
form~\cite{kitag08}
\begin{equation}
\label{hfs}
H=B\ ({\vec I}\cdot {\vec S}) +
C \ ({\vec I}\cdot {\vec N})({\vec S}\cdot {\vec N})\ ,
\end{equation}
where ${\vec N}$ is a unit vector directed from As to Fe, see 
Fig.~\ref{cell}.
The B-term is due to $s$-wave transferred hyperfine interaction
and the C-term is due to $p$-wave transferred hyperfine interaction.
It is known~\cite{kitag08} that the effective static hyperfine
magnetic field at As takes the value
 $B_\text{eff}\approx 1.5$Tesla and is directed along the $c$ axis.
The average electronic magnetic moment at the Fe site
is $\approx 0.8\mu_B$~\cite{zhao08,ewing08}. Assuming that $g=2$
this implies an average static spin component of
 $\langle S\rangle \approx 0.4$.
Since the electronic spins are arranged in a staggered pattern, see 
Fig.~\ref{cell}, only the C-term in (\ref{hfs}) contributes to the 
effective static field. Hence, 
\begin{equation}
\label{cc}
\mu_n B_{\text{eff}} = 4\times 0.4\times 0.33\times C\times I\ ,
\end{equation}
 where
$I=3/2$ is the nuclear spin and $\mu_n=1.86\mu_N$ is the 
magnetic moment of the $^{75}$As nucleus.
The factor 4 in (\ref{cc}) stems from the four neighboring Fe ions and 
$0.33=N_aN_c$ is a product of components of the unit vector
${\vec N}$ in the geometry of the As-Fe$^4$ pyramid.
From (\ref{cc}) we find
\begin{equation}
\label{ccc}
C\approx 1.1\times 10^{-4}\text{meV} \to 27 \text{MHz} \ .
\end{equation}

Alternatively, the rescaling from the known values of transferred hyperfine 
constants in cuprates yields the following estimates
\begin{subequations}
\label{BC}
\begin{eqnarray}
B &\approx& 10^{-3}\text{meV} \to 250\text{MHz}\ ,\\
C &\approx& 10^{-4}\text{meV} \to 25\text{MHz}\ .
\end{eqnarray}
\end{subequations}
While the estimate for $C$ agrees very well with (\ref{ccc})
deduced from the experimental data \cite{kitag08,baek09,curro09}, 
the estimate (\ref{BC}) for $B$ is about five times
larger than the one  measured in Ref.~\onlinecite{kitag08}.
It is worthwhile noting that there is a comment in Ref.~\onlinecite{kitag08} 
that they might underestimate the value of $B$.
In the present work, we will rely on the value of $C$ given in (\ref{ccc})
and on the estimate for $B$ given in (\ref{BC}).
It is very natural that $B \gg C$ because $B$ is due to the $s$-wave and $C$ 
is due to the $p$-wave hyperfine interaction.

The static components of the electron spins are polarized along the $a$-axis,
see Fig.\ \ref{cell}. Hence the spin wave excitations are 
polarized along the $b$- and the $c$-axis. 
To describe the magnon Raman process  shown in Fig.\ \ref{hfsfig} we need
only the part of (\ref{hfs}) that is bilinear in spin wave
creation and annihilation operators, i.e., bilinear in $\vec{\pi}$
in the language of the non-linear $\sigma$ model.
This implies that we only need to keep the terms in (\ref{hfs}) that are 
proportional to the $a$-component of the electron spin $S$
\begin{equation}
\label{hfs1}
H\to \left(B\ I_a  + 0.33C  I_c\right) \sum_iS_{ai} \ ,
\end{equation}
where the summation goes over four nearest Fe sites.
Finally, in the notation of the $\sigma$-model, 
${\vec S}/S_\text{eff} \to {\vec n}
=(n_a,n_b,n_c)=(\sqrt{1-\pi^2},  \vec{\pi})$,
this leads to
\begin{eqnarray}
\label{hfs3}
H&\to& \delta({\bf r})S_\text{eff}
(2\ B\ I_a \ \partial_a + 4 \ 0.33C \ I_c) n_a\nonumber\\
&\to&
\delta({\bf r})S_\text{eff}
(B\ I_a \ \partial_a  + 0.66C \ I_c) {\vec \pi}^2 \ ,
\end{eqnarray}
where constant terms are omitted in passing to the last line.
The gradient $\partial_a$ along the $a$-axis in the 
$B$-term appears because 
the magnetization in this direction is staggered.
We remind the reader that in our notations both $\delta({\bf r})$ and 
$\partial$ are dimensionless.
In Eq.(\ref{hfs3}) we have introduced the effective spin $S_\text{eff}$.
The first naive impression is that $S_\text{eff}=\langle S\rangle \approx 0.4$.
This would imply that the magnon Raman operator is renormalized 
by quantum fluctuations exactly like the staggered magnetization.
However, we have checked by an explicit single loop calculation 
that the Raman operator is not renormalized while the staggered magnetization
is certainly reduced in the single loop approximation.
So the naive expectation is wrong.
For numerical estimates we will use
\begin{equation}
\label{seff}
S_\text{eff}=1 \ .
\end{equation}

It is clear from the kinematic structure of Eq.\ (\ref{hfs3}) that the $B$-term
contributes to the spin-lattice relaxation only if the initial nuclear spin
is directed perpendicular to the $a$-axis. The $C$-term contributes to the
relaxation only if the initial nuclear spin is directed perpendicular to the
$c$-axis. The  Raman relaxation rate due to the $B$-term
is given by  Fermi's Golden Rule
\begin{eqnarray}
\label{WB}
W_B &=& 2\pi \frac{[ S_\text{eff} B]^2}{V^2} \\
&\times&\sum_{{\bf k},{\bf q}}
\frac{(k_a-q_a)^2}{\chi_{\perp}^2\omega_k\omega_q}n_k(1+n_q)
\delta(\omega_k-\omega_q-\omega_\text{NMR}) \ .\nonumber
\end{eqnarray}
The factor $2\pi=2\times \frac{1}{2} \times 2\pi$ is the factor $2\pi$ from
Fermi's Golden Rule  multiplied by the number of magnon polarizations, 2,
and multiplied by $\frac{1}{2}$ resulting the from
Clebsch-Gordon coefficients related to the nuclear spin $I=3/2$.
The factor $(\chi_{\perp}^2\omega_k\omega_q)^{-1}$
is due to the normalization of the ${\vec \pi}$ field, the factor
$(k_a-q_a)^2$ is due to the gradient $\partial_a$, and $n_q$ is given 
by (\ref{BE}).
Since the NMR frequency is very small, $\omega_\text{NMR} \ll \omega_q$,
the expression (\ref{WB}) can be transformed to
\begin{equation}
\label{WB1}
W_B = \frac{[ S_\text{eff} B]^2}{12\pi^3(\chi_{\perp}v_av_bv_c)^2}
\frac{1}{v_a^2}\int_{\Delta}^{\infty}\frac{(\omega^2-\Delta^2)^2}
{\sinh^2(\omega/2T)}d\omega \ ,
\end{equation}
where $\Delta$ is the spin wave gap. The two factors $(\omega^2-\Delta^2)$
in the numerator of the integrand stem from the density-of-states
in three dimensions and from the matrix element $(k_a-q_a)^2$.

Similarly, the  Raman relaxation rate due to the $C$-term reads
\begin{equation}
\label{WC1}
W_C = \frac{[ S_\text{eff} 0.66C]^2}{8\pi^3(\chi_{\perp}v_av_bv_c)^2}
\int_{\Delta}^{\infty}\frac{(\omega^2-\Delta^2)}
{\sinh^2(\omega/2T)}d\omega \ ,
\end{equation}
where there is one factor $(\omega^2-\Delta^2)$ less
 in the numerator because there is no particular momentum dependent
matrix element.

\begin{figure}[ht]
\includegraphics[width=0.98\columnwidth,clip]{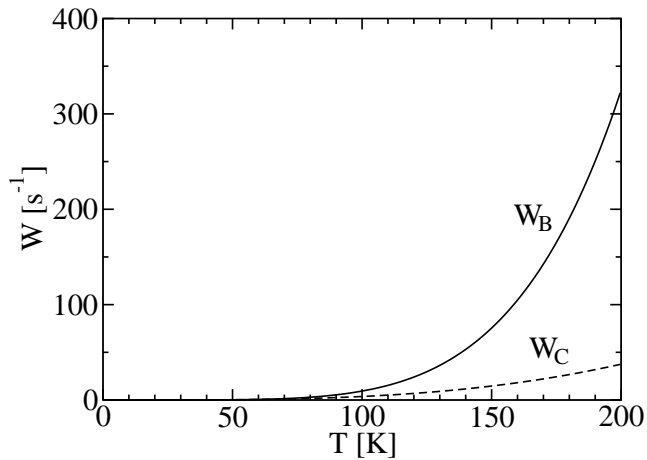}
\caption{
NMR relaxation rates due to magnon scattering.
The solid curve gives contribution of the $s$-wave transferred
hyperfine coupling, see Eq.\ \eqref{WB1}.
The dashed curve gives contribution of the $p$-wave transferred
hyperfine coupling, see Eq.\ \eqref{WC1}.
The parameters are given by (\ref{vac}), 
(\ref{chi2}), (\ref{BC}), (\ref{seff}) and $v_b=20$meV. 
}
\label{FWB}
\end{figure}

It is clear that Eqs.\ (\ref{WB1}) and (\ref{WC1}) are not justified in the
 vicinity of the N\'eel temperature. Obviously, they  are not valid at
$T > T_N$ either. So we use them only below $T_N$.
Both $W_B$ and $W_C$ are very steep functions of temperature.
Plots of $W_B$ and $W_C$ calculated with parameters given by (\ref{vac}), 
(\ref{chi2}), (\ref{BC}), (\ref{seff}) and $v_b=20$meV are presented in 
Fig.~\ref{FWB}.
%%%%%%%%%%%%%%%%%
%%%%%%%%%%%%%%%%%
The decay rate $W_C$ is much smaller than $W_B$, 
$W_B \gg W_C$, and hence $1/T_1\approx W_B$
so that the magnon Raman relaxation is mainly due to the $s$-wave
transferred hyperfine interaction.
The estimate for the value of relaxation rate presented in Fig.~\ref{FWB}
agrees with the available data~\cite{kitag08,baek09,curro09}.

One can fit the data by fine tuning of $B$ and/or $v_b$ and/or $\chi_{\perp}$.
But this is not our aim here.
Note, however, that the NMR relaxation rate (\ref{WB1}) is
proportional to the second power of the density of magnetic 
excitations (\ref{md3}). So it is very sensitive to the value
of $\rho$ in (\ref{md3}) and hence to the value of $v_b$.
Hence the fact that our previous set 
of parameters yields the correct magnitude of the NMR relaxation rate
strongly supports our claim that $v_b$ is small.

Since the relaxation is dominated by $W_B$ we  predict
a significant polarization dependence of the relaxation.
The corresponding Hamiltonian contains only the $a$-component of the nuclear
spin, see Eqs.\ (\ref{hfs1}) and (\ref{hfs3}).
Therefore, this mechanism does not contribute to relaxation
if the $^{75}$As nuclear spin is polarized along the $a$-axis.
The mechanism contributes equally for polarizations along the $b$- and 
the $c$-axis.
For a twinned sample, where domains with swapped $a$ and $b$ axes are
of equal weight,  this argument implies that the relaxation for the 
$c$-polarization of nuclear spin is twice faster than the relaxation 
for an in-plane polarization of the nuclear spin.

For temperatures below the spin-wave gap the magnon Raman relaxation
is essentially switched off. In other words, 
the mechanism related to collective magnetic modes is not active.
But there is also a diffuse magnetic relaxation stemming from the
charge degrees of freedom because the system is not an insulator.
It is natural to assume that this diffuse relaxation scales linearly
with temperature as it does in normal Fermi liquids. 
This low-temperature behavior of the relaxation was observed in  
Refs.~\onlinecite{kitag08} and \onlinecite{curro09}. The charge 
driven relaxation
 is certainly also active for $T > \Delta$. But in this regime its
contribution to the NMR relaxation rate is relatively small with the
contribution from collective magnetic modes prevailing.\\

\section{Conclusions}

We studied the parent ferro-pnictides below their N\'eel temperature.
Based on the nonlinear $\sigma$-model, we considered
 the available experimental data on the temperature
dependence of the staggered magnetization phenomenologically.
We found that one needs a high density of magnetic excitations
to explain the relatively strong temperature dependence of the magnetization.
This implies that the spin wave velocity along the spin stripes
is very small. The values for this velocity estimated from the analysis are
$v_b\approx 10-30$meV. For comparison, the in-plane velocity perpendicular
to stripes takes the value $v_a \approx 205$meV.

We also analyzed the NMR spin-lattice relaxation rate for  $^{75}$As.
Due to their high spectral density  the magnons dominate 
the relaxation rate at temperatures above the spin-wave gap.
Our estimates for the relaxation rate based
on the density found from the neutron scattering data agree very well
 with direct NMR measurements.
This is an independent confirmation of the high spectral density of magnetic
excitations.

So both the temperature dependent magnetization as well as the NMR
relaxation rate confirm strong magnetic fluctuations at low
energies. Thus the present phenomenological analysis corroborates
the scenario that the ferro-pnictides constitute systems close
to a quantum phase transition triggered by frustrated magnetic
couplings \cite{uhrig09a}. Hence a strongly correlated
picture of the ferro-pnictides is favored.

Very recent inelastic neutron scattering data~\cite{diallo09,zhao09}
 indicate the ratio of spin wave velocities $v_a/v_b \approx 2$.
This is not consistent with our conclusion $v_a/v_b \approx 10$.
Our analysis of the temperature dependence
of the staggered magnetization and especially of the NMR relaxation
rate is in essence based only on the spin-wave dispersion (\ref{disp}).
Only the dispersion determines the density of excited magnons at a
given temperature, and only the density determines the NMR
relaxation rate. With the spin-wave velocity $v_b$ taken 
from~\cite{diallo09,zhao09} one obtains the relaxation rate about 20-30
smaller than the experimental one.

 How can the above discrepancy be explained?
The data \cite{diallo09,zhao09} is taken on twinned samples
because only below the structural transition temperature
the orthorhombicity occurs.
Superposing dispersions with prominent ridges such as the ones in Fig.\
3a in Ref.\ \onlinecite{uhrig09a} can lead to responses similar
to the ones in Ref.\ \cite{diallo09,zhao09} for moderate and
high energies. The time-of-flight 
technique used in both experimental probes \cite{diallo09,zhao09}
is certainly best
suited for investigating the moderate and higher energies.

If the careful study of the influence of twinning does not solve the 
discrepancy our analysis indicates the existence of some
low-energy ($\approx 10-20$meV) magnetic degrees of freedom
which have so far not been taken into account.
These degrees of freedom must contribute to the NMR relaxation and they
must be difficult to detect by neutron scattering.

From our results and the above discussion we conclude that
further experiments focusing on low lying magnetic modes
are called for to resolve this crucial issue. It would
be highly desirable if low-temperature detwinned samples could be generated.

\acknowledgments
We thank  A.~A.~Katanin, N.J.~Curro, G.~Khaliullin,
A.I.~Milstein, R.R.P.~Singh, and C.Ulrich for helpful discussions
and J.~Zhao for providing the experimental data in Fig.\ \ref{exp}.
G.S. Uhrig acknowledges financial support by the Heinrich-Hertz Stiftung
NRW and the Gordon Godrey Fund.

%\bibliographystyle{apsrev}
%\bibliography{../../bibinput/liter10}

\end{document}